\magnification=1200 \vsize=25truecm \hsize=16truecm \baselineskip=0.6truecm
\parindent=1truecm \nopagenumbers \font\scap=cmcsc10 \hfuzz=0.8truecm
\font\tenmsb=msbm10
\font\sevenmsb=msbm7
\font\fivemsb=msbm5
\newfam\msbfam
\textfont\msbfam=\tenmsb
\scriptfont\msbfam=\sevenmsb
\scriptscriptfont\msbfam=\fivemsb

\def\yup{\overline y}
\def\xup{\overline x}
\def\pup{\overline p}
\def\pdo{\underline p}
\def\pdd{\underline{\underline{p}}}

\def\cdo{\underline c}
\def\xdo{\underline x}
\def\cdd{\underline{\underline{c}}}

\null \bigskip  \centerline{\bf SCHLESINGER TRANSFORMATIONS FOR
LINEARISABLE EQUATIONS}

\vskip 2truecm
\bigskip
\centerline{\scap A. Ramani}
\centerline{\sl CPT, Ecole Polytechnique}
\centerline{\sl CNRS, UPR 14}
\centerline{\sl 91128 Palaiseau, France}
\bigskip
\centerline{\scap B. Grammaticos}
\centerline{\sl GMPIB, Universit\'e Paris VII}
\centerline{\sl Tour 24-14, 5$^e$\'etage}
\centerline{\sl 75251 Paris, France}
\bigskip
\centerline{\scap S. Lafortune$^{\dag}$}
\centerline{\sl LPTM et GMPIB,  Universit\'e Paris VII}
\centerline{\sl Tour 24-14, 5$^e$\'etage}
\centerline{\sl 75251 Paris, France}
\footline{\sl $^{\dag}$ Permanent address: CRM, Universit\'e de
Montr\'eal, Montr\'eal, H3C 3J7 Canada}
\bigskip\bigskip

Abstract
\smallskip \noindent
We introduce the Schlesinger transformations of the Gambier equation. The
latter can be written, in both the continuous
and discrete cases, as a system of two coupled Riccati equations in cascade
involving an integer parameter $n$. In the
continuous case the parameter appears explicitly in the equation while in
the discrete case it corresponds to the number
of steps for singularity confinement. Two Schlesinger transformations are
obtained relating the solutions for some value
$n$ to that corresponding to either $n+1$ or $n+2$.
\vfill\eject

\footline={\hfill\folio} \pageno=2

\bigskip
\noindent {\scap 1. Introduction}
\medskip

The existence of Schlesinger transformations is one of the very special
properties of Painlev\'e
equations [1]. These transformations are a particular kind of
auto-B\"acklund transformations [2]. The latter
relate a solution of a given equation to a solution of the same equation
but corresponding to a
different set of parameters. Schlesinger transformations do just that but
the changes of parameters
correspond to integer or half-integer shifts in the monodromy exponents.
Both continuous and discrete
(whether difference or multiplicative) Painlev\'e equations have been shown
to possess Schlesinger
transformations [3]. For the discrete case (and in particular for
$q$-Painlev\'e's) the relation of
Schlesinger transformations to monodromy exponents is not quite clear and
their derivation requires
both experience and intuition, in particular in the choice of the proper
parameters. With this minor
{\sl caveat} we hasten to say that we do possess a systematic approach to
the construction of Schlesinger
transformations. It is based on the bilinear formalism [4] which can be
used to construct Miura
transformations [5], the iteration of which can lead to auto-B\"acklund's.

The logical conclusion of the above introduction is that Schlesinger
transformations should exist
only for Painlev\'e equations. This is almost true with one exception. As
we shall report in this paper there
exists one equation which, without being a Painlev\'e, does possess
Schlesinger's. This equation is known
under the name of Gambier (who first derived it): it is the most general
second-order ODE of
linearisable type [6]. What makes possible the existence of Schlesinger
transformations for this equation
is the fact that its general expression involves an arbitrary integer $n$.
It turns out that we can
relate the solution of the equation for some value of $n$ to that
corresponding to $n+2$. These
transformations are, for the linearisable case, the analogues of the
Schlesinger's. Moreover, as we shall show,
the same procedure can be followed in the discrete case i.e. for the
Gambier mapping [7,8].

In section 2 we shall review some basic facts about the Gambier equation.
The Schlesinger
transformations will be given in section 3 while section 4 is devoted to
the study of the discrete case.

\bigskip
\noindent {\scap 2. The Gambier equation}
\medskip

The Gambier equation is given as a system of two Riccati equations in
cascade. This means that
we start with a first Riccati for some variable $y$
$$
y'=-y^2+by+c \eqno(2.1)
$$
and then couple its solution to a second Riccati by making the coefficients
of the latter depend
explicitly on $y$:
$$
x'=ax^2+nxy+\sigma. \eqno(2.2)
$$

The precise form of the coupling introduced in (2.2) is due to
integrability requirements. In fact, the
application of singularity analysis shows that the Gambier system cannot be
integrable unless the
coefficient of the $xy$ term in (2.2) is an integer $n$. This is not the
only integrability
requirement. Depending on the value of $n$ one can find constraints on the
$a$, $b$, $c$, $\sigma$
(where the latter is traditionnally taken to be constant $1$ or $0$) which
are necessary for
integrability.

The common lore [9] is that out of the functions $a$, $b$, $c$ two are
free. This turns out not to be
the case. The reason for this is that the system (2.1-2) is not exactly
canonical i.e. we have not used
all possible transformations in order to reduce its form. We introduce a
change of independent
variable from $t$ to $T$ through
$dt=gdT$ where
$g$ is given by ${1\over g}{dg \over dt}=b{n\over 2-n}$, a gauge through
$x=gX$ and also
$Y=gy-{1\over n}{dg\over dt}$.
The net result is that system (2.1-2) reduces to one where $b=0$ while
$\sigma$ remains equal to $0$ or $1$. It is clear
from the equations above that $n$ must be different from $2$. On the other
hand when $n=2$ the integrability condition,
if $\sigma=1$, is precisely
$b=0$. So we can always take $b=0$. (As a matter of fact in the case
$\sigma=0$ an additional gauge
freedom allows us to take both $b$ and $c$ to zero for all $n$, even for
$n=2$). Thus the Gambier system can be written in
full generality
$$
y'=-y^2+c
\eqno(2.3a)
$$
$$
x'=ax^2+nxy+\sigma.
\eqno(2.3b)
$$

One further remark is in order here. The system (2.3) retains its form
under the transformation
$x\rightarrow 1/x$. In this case $n \rightarrow -n$ and $\sigma$ and $-a$
are exchanged. Thus in
some cases it will be interesting to consider a Gambier system where
$\sigma$ is not constant but rather a
function of $t$. Still, it is possible to show that we can always reduce
this case to one where
$\sigma=1$, while preserving the form of (2.3a) i.e. $b=0$. To this end we
introduce the change of
variables $dt=hdT$, $x=gX$ and $Y=hy-{1\over 2}{dh\over dt}$ with
$h=\sigma^{2/(n-2)}$,
$g=\sigma^{n/(n-2)}$. With these transformations system (2.3) reduces to
one with $\sigma=1$ and $b=0$. (In the special
case $n=2$, with $b=0$ integrability implies $\sigma=$constant, whereupon
its value can always be reduced to $1$).

\bigskip
\noindent {\scap 3. Schlesinger transformations for the Gambier equations}
\medskip

The theory of auto-B\"acklund transformations of Painlev\'e equations is
well established. As was shown
in [2] the general form of auto-B\"acklund transformations for most
Painlev\'e equations is of the form:
$$
\tilde{x}={\alpha x'+\beta x^2+\gamma x+\delta \over \epsilon x'+\zeta
x^2+\eta x+\theta }.
\eqno(3.1)
$$
In the case of the Gambier equation considered as a coupled system of two
Riccati's it is more
convenient to look for an auto-B\"acklund of the form:
$$
\tilde{x}={\alpha xy+\beta x+\gamma y+\delta \over  (\zeta y+\eta)(\theta x
+\kappa)}.
\eqno(3.2)
$$
with a factorized denominator, with hindsight from the discrete case.
We require that the equation satisfied by $\tilde x$ do not comprise terms
nonlinear in $y$. We examine first
the case $\zeta\neq 0$ and reach easily the conclusion that there exists no
solution. So we take $\zeta
=0$, $\eta=1$ which implies that $\alpha$ and $\gamma$ do not both vanish
(otherwise (3.2) would have been independent of $y$). We find in this case
$\alpha=0$ and thus the general form of the
auto-B\"acklund can be written as:
$$
\tilde{x}={\beta x+\gamma y+\delta \over \theta x+\kappa}. \eqno(3.3)
$$
From (3.3) we can obtain the two possible forms of the Gambier system
auto-B\"acklund:
$$
\tilde{x}=\beta x+\gamma y+\delta \eqno(3.4)
$$
$$
\tilde{x}={\beta x+\gamma y+\delta \over x+\kappa}. \eqno(3.5)
$$
 As we shall
see in what follows both forms lead to Schlesinger transformations.

Let us first work with form (3.4). Our approach is straightforward. We
assume (3.4) and require that
$\tilde x$ satisfy an equation of the form (2.3b) while $y$ is always the
same solution of (2.3a).
The calculation is easily performed leading to:
$$
\tilde{x}=\gamma y+{a \gamma \over n+1}x+{\gamma' \over n}, \eqno(3.6)
$$
where $\gamma$ satisfies:
$$
{\gamma' \over \gamma}={n \over n+2}{a' \over a}. \eqno(3.7)
$$
Here we have assumed $a\neq 0$ otherwise $\tilde x$ does not depend on $x$
and (3.6) does not define a Schlesinger. The
parameters of the equation satisfied by
$\tilde x$ are given (in obvious notations) by:
$$
\tilde{n}+n+2=0 \eqno(3.8a)
$$
$$
\tilde{a}={n+1 \over \gamma} \eqno(3.8b)
$$
and
$$
\tilde{\sigma}=\gamma\Big(c+{a\sigma \over n+1}+{1\over n+2}{a''\over
a}-{n+3\over (n+2)^2}{a'^2\over
a^2}\Big).
\eqno(3.8c)
$$
Thus (3.6) is indeed a Schlesinger transformation since it takes us from a
Gambier system with parameter
$n$ to one with parameter $\tilde n=-n-2$. It suffices now to invert
$\tilde x$ in order to obtain an equation
with parameter $N=n+2$. Expressions (3.6) and (3.8) can be written in a
more symmetric way:
$$
\tilde{a}\tilde{x}-a x=(n+1)(y-{a' \over \tilde{n}a}) \eqno(3.9)
$$
and
$$
\matrix{
\displaystyle{\tilde{n}+1=-(n+1)}\cr
\displaystyle{\tilde{n}{\tilde{a}' \over \tilde{a}}=n{a' \over a}}\cr
\displaystyle{\tilde{a}\tilde{\sigma}-a\sigma=(n+1)\Big(c-{1\over \tilde{n}}
\big({a'\over a}\big)'+{1\over\tilde{n}^2}{a'^2\over a^2}\Big). }
}
\eqno(3.10)
$$

The inverse transformation can be easily obtained if we introduce $\tilde
\gamma$ such that
$a\tilde{\gamma}=-(n+1)=-\tilde{a}\gamma$. We thus find
$$
x=y \tilde{\gamma}+{\tilde{a}\tilde{\gamma}\over
\tilde{n}+1}\tilde{x}+{\tilde{\gamma}' \over \tilde{n}}
\eqno(3.11)
$$
and the relations (3.10) are still valid.

Iterating the Schlesinger transformations one can construct the integrable
Gambier systems for higher
$n$'s and obtain by construction the functions which appear in them.
However it may happen that when we implement the Schlesinger we find
$\tilde{\sigma}=0$. If we invert $x$ we get a system with
$N=-\tilde{n}=n+2$ but $A=0$ for which one cannot iterate the
Schlesinger further.

Let us give example of the application of this Schlesinger transformation.
Let us start from $n=0$,
in which case we find $\tilde{n}=-2$ and, after inversion, $N=2$. For $n=0$
we start from $a=-1$ and
$\sigma=0$ or $1$ (always possible through the appropriate changes of
variable). This leads to
$\tilde{a}=-1$, $\tilde{\sigma}=-c+\sigma$ and the Schlesinger reads:
$\tilde{x}=-y+x$. Next we invert
$\tilde{x}$ and have $X=1/(x-y)$. We find thus that the Schlesinger takes
us from
$$
\matrix{
y'=-y^2+c \cr
x'=-x^2+\sigma
}
\eqno(3.12)
$$
to the system
$$
\matrix{
y'=-y^2+c \cr
X'=AX^2+2Xy+\Sigma
}
\eqno(3.13)
$$
with $A=c-\sigma$, $\Sigma=1$. In the particular case $n=2$, a change of
variables exists which allows us to put $A=-1$
(unless $A=0$), without introducing $b$ in the equation for $y$, while
keeping $\Sigma=1$ and changing only the value of
$c$. Thus the generic case of the Gambier equation for $n=2$ can be written
with $A=-1$. Eliminating
$y$ between the two equations we find:
$$
x''={x'^2 \over 2x}-2xx'-{x^3 \over 2}-{1 \over 2x}+(2c+1)x. \eqno(3.14)
$$
This is the generic form of the $n=2$ Gambier equation and it contains just
one free function. The nongeneric cases
corresponding to $A=0$ and $\sigma=0$ or $1$ can be constructed in an
analogous way.

We now turn to the second Schlesinger transformation corresponding to the
form (3.5). As we shall show,
a Schlesinger transformation of this form does indeed exist and corresponds
to changes in $n$ with $\Delta
n=1$. Let us start from the basic equations (2.3). Next we ask that $\tilde
x$ defined by (3.5) indeed
satisfy a system like (2.3). We find thus that and $\kappa=-x_0$ and
$\gamma$ must be given by:
$$
{\gamma' \over \gamma}=y_0+{2ax_0 \over n+1} \eqno(3.15)
$$
where
$y_0$ is a solution of the Riccati (2.3a)
and  $x_0$ a solution of (2.3b), obtained with $y$ replaced by $y_0$. We
introduce the quantities $\tilde{x}_0={a\gamma
\over n+1}$,
$\tilde{a}=-{n x_0\over \gamma}$. In this case (3.15) becomes:
$$
{\gamma' \over \gamma}=y_0+{2\tilde{a}\tilde{x}_0 \over
\tilde{n}+1}=y_0+{2x_0\tilde{x}_0 \over \gamma},
\eqno(3.16)
$$
where
$$
\tilde{n}+n+1=0. \eqno(3.17)
$$
We have thus, starting from a generic solution $x$, $y$ of (2.3) for some
$n$, the Schlesinger:
$$
\tilde{x}=\tilde{x}_0+{\gamma(y-y_0) \over x-x_0} \eqno(3.18)
$$
where $\tilde{x}$  is indeed a solution of (2.3) for $\tilde{n}=-n-1$ for
the {\sl same} $y$
$$\tilde{x}'=-\tilde{a}\tilde{x}^2+\tilde{n}\tilde{x}y+\tilde{\sigma}\eqno(3.19)
$$
where $\tilde{a}$ has been defined as  $-{n x_0\over \gamma}$ and
$$\tilde{\sigma}={\gamma \over n+1}\Big(a'+a^2x_0{n+2\over
n+1}+ay_0(n+2)\Big). \eqno(3.20)$$
Note that $\tilde{x}_0$ is a solution of the same equation with $y$
replaced by $y_0$.
As in the previous case if we invert $\tilde x$ we obtain an equation
corresponding to $N=n+1$.

As an application of the $\Delta n=1$ Schlesinger we are going to construct
the $n=1$ equation starting from the $n=0$
case, i.e. system (2.3) with $n=0$. From (3.18), the Schlesinger reads
$$
\tilde{x}=\gamma\big(a+{y-y_0 \over x-x_0}\big)\eqno(3.21)
$$
where $\gamma$ satisfies the differential equation
$$
{\gamma' \over \gamma}=y_0+2ax_0 \eqno(3.22)
$$
and where $y_0$ is a particular of (2.3b) and $x_0$ is a solution of (2.3a)
with $n=0$. Then  $\tilde x$
satisfies a Gambier equation in which $n=-1$, $\tilde{\sigma}=\gamma(a'+2
a^2x_0+2ay_0)$ and $\tilde{a}=0$. To obtain
the
$n=1$ Gambier equation, we define
$X=1/\tilde{x}$ and we arrive at the following system
$$
\matrix{
\displaystyle{y'=-y^2+c} \cr \cr
\displaystyle{X'=AX^2+Xy}
}
\eqno(3.23)
$$
where $A=-\tilde \sigma$. The system (3.23) is the generic $n=1$ case
since, in this case, the condition for (2.3) to
be integrable is $\sigma=0$.

It is worth pointing out here that the Schlesinger transformation
corresponding to $\Delta n=2$ was known
to Gambier himself. As a matter of fact when faced with the problem of
determining the functions
appearing in his equation so as to satisfy the integrability requirement,
Gambier proposed a recursive
method which is essentially the Schlesinger $\Delta n=2$. On the other hand the
Schlesinger
$\Delta n=1$ is quite new and we have first discovered it in the discrete
case whereupon we looked for
(and found) its continuous analogue.

\filbreak
\bigskip
\noindent {\scap 4. The Gambier mapping}
\medskip
The Gambier equation has been examined already in [7,8] and its discrete
equivalent has been proposed
there. These constructions of the Gambier mapping were {\sl ad hoc} ones,
in the sense that we assumed a
form and implemented the singularity confinement discrete integrability
criterion in order to obtain
the integrability conditions. In what follows, we shall use a slightly
different approach based on the
singularity structure.

Our starting point is the discrete equivalent of the system (2.3). We have
thus one equation
which is the discrete analogue of the Riccati, i.e. a homographic mapping
for $y$ and another
homographic mapping for $x$, the coefficients of which depend linearly on
$y$. Our derivation will be
based on the study of singularities of the system. The general homographic
equation for $y$ involves
three free parameters, but since we have the freedom of choice of a
homographic transformation
on $y$, we can always reduce it to $y=$constant (i.e. $\yup=y$). However,
in the system under study,
our aim is to study the singularities of $x$ induced by {\sl special}
values of $y$. One could choose the
singularity to enter at point $n_0$ if the value of $y$ has some special
value depending on $n_0$, say,
$f(n_0)$. This would introduce one function in the homographic mapping.
However what is even more convenient is to decide what the special value of
$y$ is, say
$y=0$ for all $n$, at the price of the loss of part of the  homographic
freedom. Then the special value $0$ will
occur for some $n$ depending on the particular solution. Of course, if we
allow the full homographic freedom we are back
to the starting point i.e. with three free functions. However we decide to
have only one free function and thus we
simplify the mapping by chosing its form so that it presents the pattern
$\{-1, 0, 1\}$. This fixes two of the functions
and the result is:
$$
\yup={y+c \over y+1} \eqno(4.1)
$$
where $c$ is a function of $n$ and we use the notations $y=y(n)$,
$\yup=y(n+1)$.

Next, we turn to the equation for $x$. This equation is homographic in $x$.
However we require that when
$y$ takes the value $0$, the resulting value of $x$ be $\infty$. Thus the
denominator must be
proportional to $y$, and since we can freely translate $x$ we can reduce
its form to just $xy$. The
remaining overall gauge factor is chosen so as to put the coefficient of
$xy$ of the numerator to unity
resulting to the following mapping:
$$
\xup={x(y-r)+q(y-s) \over xy}. \eqno(4.2)
$$
The system (4.1-2) is a discrete form of the Gambier system. In order to
study the confinement of the
singularity induced by $y=0$ we introduce the auxiliary quantity $\psi_N$
which is the $N$'th iterate of
$y=0$ in equation (4.1), $N$ times downshifted. Thus $\psi_0=0$,
$\psi_1=\underline{c}$,
$\psi_2={\underline{\underline{c}}+\underline{c} \over
\underline{\underline{c}} +1}$, etc... The
confinement requirement is that after $N$ steps $x$ becomes $0$ in such a
way as to lead to $0/0$ at the
next step. Thus the mapping (4.2) has in fact the form:
$$
\xup={x(y-r)+q(y-\psi_N) \over xy}. \eqno(4.3)
$$
Thus when at some step $N$ we have $y=\psi_N$ and $x=0$. On the view of
(4.3)  $\xup$ will then be indeterminate of the
form $0/0$. However it turns out that, in fact, this value is
well-determined and finite. Let us take a closer look at the
conditions for confinement. The generic patterns for $x$ and $y$ are:
$$
\matrix{
\displaystyle{ y: \{} & & \displaystyle{0}&& \displaystyle{\overline{\psi_1}}&&
\displaystyle{\overline{\overline{\psi_2}}}&& \displaystyle{\dots}
&&\displaystyle{\overline{\dot{\dot{\overline{\overline{{\psi}}}}}}_N}&&
\} \cr
\displaystyle{x: \{}& {\rm free}&& \infty &&
\displaystyle{{\overline{\psi_1}-\overline{r} \over
\overline{\psi_1}}}&&\dots && 0
&&{\rm free}& \ \}\hfill. }
$$
At $N=1$ it is clearly impossible to confine with a form (4.3) since we do
not have enough steps. In
this case the only integrable form of the $x$-equation is a linear one. The
first genuinely confining
case of the form (4.3) is $N=2$. From the requirement $\overline{\xup}=0$
we have $r=\psi_1$ and $q$ free: this is
indeed the only integrability condition. For higher $N$'s we can similarly
obtain the confinement
condition which takes the form of an equation for $r$ in terms of $q$.

At this point it is natural to ask whether the mapping (4.1)-(4.3) does
indeed correspond to the Gambier equation
(2.3). In order to do this we construct its continuous limit. We first
introduce:
$$
\matrix{
\displaystyle{c=\epsilon^2D} \cr \cr
\displaystyle{y={\epsilon D \over Y+H}}
}
\eqno(4.4)
$$
with $H\approx D'/(2D)$ and obtain the continuous limit of (4.1) for
$\epsilon \to 0$. We find as
expected
$$
Y'=-Y^2+C \eqno(4.5)
$$
i.e. eq. (2.3a), where
$C=D-{D''\over 2D}+{3\over 4}{D'^2 \over D^2}$. Using (4.4) and (4.1) we
can also compute $\psi_N$ and we find at lowest
order:
$$
\psi_N=\epsilon^2\Psi_N
\qquad{\rm
with
}\qquad
\Psi_N\approx N(D-\epsilon{N+1 \over 2}D')+\epsilon^2\Phi_N. \eqno(4.6)
$$
where $\Phi_N$ is an explicit function of $D$ depending on $N$.

Next we turn to the equation for $x$ and introduce:
$$
\matrix{
\displaystyle{r=\epsilon^2 R} \cr \cr
\displaystyle{x\approx{1\over 2}+{\epsilon \over 2X}-\epsilon {R D' \over
4D^2}} \cr \cr
\displaystyle{q\approx-{1\over 4}+\epsilon^2Q}
}
\eqno(4.7)
$$
and for the continuous limit of the form (2.3b) to exist in canonical form
(i.e. $b$=0, $\sigma$=1) we find that we must
have
$$
R\approx{ND\over 2} -\epsilon (N+2){ND' \over 8}.
\eqno(4.8)
$$
This leads to the equation for $x$:
$$
X'=AX^2+NXY+1 \eqno(4.9)
$$
with $A={N\over 4}(N/4+1){D'^2 \over D^2}-{ND'' \over 4D}-4Q$.
Moreover the confinement constraint implies a differential relation between
$D$ and $Q$ which depends on $N$. We can
verify explicitly in the first few cases that this is indeed the
integrability constraint obtained in the continuous
case. For instance, for $N=2$, just imposing (4.8) in order to have the
canonical form $b$=0, $\sigma$=1, is sufficient for
integrability.

 Once the singularity pattern of the Gambier mapping is established we can
use it in order to construct the
Schlesinger transformation. Let us first look for a transformation that
corresponds to
$\Delta N=2$. The idea is that given the $N$-steps singularity pattern of
the equation for $x$ we introduce a
variable $w$ with $N+2$ singularity steps where we enter the singularity
one step before $x$ and exit
it one step later. The general form of the Schlesinger transformation,
which defines $w$, is:
$$
w=X{y-\psi_{N+1} \over y}, \eqno(4.10)
$$
where $X$ is homographic in $x$. The presence of the $y$ and $y-\psi_{N+1}$
terms is clear: they ensure
that $w$ becomes infinite one step before $x$, and vanishes one step after
$x$. Next we turn to the
determination of $X$. Since $X$ is homographic in $x$ we can rewrite (4.10) as:
$$
w={\alpha x+\beta \over y}{y-\psi_{N+1} \over \gamma x+\delta}. \eqno(4.11)
$$
 Our requirement is that $w$  becomes infinite when $y=0$ for every value
of $x$. This statement
must be qualified. The numerator $\alpha x+\beta$ {\it will} vanish for
some $x$ (namely
$x=-\beta/\alpha$) so this value of $x$ must be the only one which should
{\it not} occur in the
confined singularity. Indeed there is a unique value of $x$ where instead
of being confined, the
singularity extends to
infinity in {\it both} directions of the independent variable $n$, while
the only nonsingular values of
the dependent variable occur in a finite range. The value of $x$ such that
$\xup$ is finite and free even
though $y$ is zero is such that the numerator $-xr-q\psi_{N}$ of  $\xup$
vanishes.
For this value of $x$, the values of the dependent
variable are fixed for $n\le 0$  and $n\ge N+1$ and the value can be
considered as `forbidden'. Thus $\alpha x
+\beta=xr+q\psi_{N}$ up to a multiplicative constant. Similarly when
$y=\psi_{N+1}$, $w$ must
vanish. Thus
$\gamma x+\delta$ must not be zero except for the unique value of $x$ that
does not occur in
the confined singularity. Note that $y=\psi_{N+1}$ means $\underline
y=\underline {\psi}_N$ and
the only value of $x$ that comes from a nonzero $\xdo$ in that case is
$x={(\underline{\psi}_N-\underline r) /\underline{\psi}_N}$. In that case
the values of the
dependent variable are fixed for $n\ge 0$ and $n\le -N-1$. This value of
$x$ being `forbidden',
$\gamma x+\delta$ must be proportional to
$\underline{\psi}_N x-(\underline{\psi}_N-\underline{r})$. We now have the
first form of the
Schlesinger:
$$
w={xr+q\psi_N \over y}{y-\psi_{N+1} \over \underline{\psi}_N x
+\underline{r}-\underline{\psi}_N}
\eqno(4.12)
$$
where the proportionality constant has been taken equal to $1$ (but any
other value would have been
equally acceptable). Here $w$ effectively depends on $x$ unless
$r(\underline{r}-\underline{\psi}_N)=q\underline{\psi}_N\psi_N$. But in
this case the mapping (4.3) is in fact linear in
the variable $\xi=(x-1+\overline{r}/\overline{\psi}_N)^{-1}$. This case is
the analog of the case $a=0$ in the
continuous case where the Schlesinger does not exist.

Let us give an application of the Schlesinger transformation by
obtaining the
$N=2$ equation starting from $N=0$. We have always the equation for $y$
which reads:
$$
\yup={ y+c \over y+1} \eqno(4.13)
$$
and $\psi_0=0$, $\psi_1=\underline{c}$.
For $N=0$ the equation for $x$ reads:
$$
\xup={x(y-r)+qy \over xy}={x+q \over x} \eqno(4.14)
$$
since for integrability $r=0$ and indeed $N=0$ means that the $x$ equation
does {\sl not} depend on $y$. We introduce the
Schlesinger:
$$
w=x{y-\underline{c} \over y} \eqno(4.15)
$$
(where we first write (4.12) for arbitrary $r$ and since $r$ factors we
take the limit $r\rightarrow 0$
afterwards). Using (4.14) and (4.15) to eliminate $x$ we obtain the
equation for $w$:
$$
\overline{w}=(1-c){yw+q(y-\underline{c}) \over (y+c)w}. \eqno(4.16)
$$
This equation is of the form (4.3) but not quite canonical. We can
transform it to canonical form simply
by introducing $\yup$ instead of $y$ because indeed $w$ is infinite one
step before $x$, so $w=\infty$ means
$\xup=\infty$ i.e. $\yup=0$. We obtain thus:
$$
\overline{w}={w(\yup-c)+q(1+\underline{c})(\yup-\overline{\psi}_2) \over
\yup w} \eqno(4.17)
$$
with $\overline{\psi}_2=(c+\underline{c})/(1+\underline{c})$ which coupled
to (4.13) is indeed a $N=2$ Gambier mapping.

As we pointed out above the $N=1$ case is not included in the
parametrisation (4.1-4.3): the $x$-mapping
must be linear in order to ensure integrability. Thus we are led to study
the linear case separately. For an arbitrary
$N$, the general form of the linear $x$-mapping can be obtained using
confinement arguments in a way similar to what
we did for the generic, nonlinear, case. We obtain
$$
\xup={x(y-\psi_N)+g \over y} \eqno(4.18)
$$
where $g$ is free. The Schlesinger transformation is again given by:
$$
w=X{y-\psi_{N+1} \over y} \eqno(4.19)
$$
and arguments similar to those of the nonlinear case allow us to determine
the form of the homographic object $X$ leading
to:
$$
w={x\psi_N-g \over y}{y-\psi_{N+1} \over
x\underline{\psi}_N-\underline{g}}. \eqno(4.20)
$$
Thus one can perform a Schlesinger in the linear case. This is not in
disagrement with the continuous case. It is, in
fact, the analog of the case where $\sigma=0$ but $a\neq 0$ (which is
linear in $1/x$) for which the Schlesinger {\sl can}
be performed. The analog of the case $\sigma=0$ and $a=0$ is the situation
when
$g=k\psi_N$ with constant
$k$ in which case  the mapping rewrites $\overline{\xi}=\xi(y-\psi_N)/y$
with $\xi=x-k$. Then $w$ does not
depend on $\xi$ (or $x$) and (4.20) does not define a Schlesinger in
analogy to the case
$r(\underline{r}-\underline{\psi}_N)=q\underline{\psi}_N\psi_N$ in the
nonlinear case.

 Using this form of the Schlesinger transformation we can, for example,
construct the $N=3$ case
starting from the $N=1$ case. In the case $N=1$, the mapping for $x$ is
given by:
$$
\xup={(y-\underline{c})x+g \over y}. \eqno(4.21)
$$
Using equation (4.20), we introduce the Schlesinger:
$$
w={x\underline{c}-g\over
y}{(\underline{\underline{c}}+1)y-\underline{c}-\underline{\underline{c}}
\over
x\underline{\underline{c}}-\underline{g}}. \eqno(4.22)
$$
In order to simplify the final expression, we define $p$ with
$g=\underline{c}p$. We then have the
following equation for $w$:
$$
\overline{w}={
c(c-1)
\Big(\,\underline{\underline{c}}
w
\big((\overline{p}-\underline{p})y+\underline{c}(\underline{p}-p)\big)
+\underline{c}(p-\overline{p})
\big(y(\underline{\underline{c}}+1)-
(\underline{\underline{c}}+\underline{c})\big)\,\Big)
\over
(p-\underline{p})\underline{c}\,\underline{\underline{c}}
w(y+c)}.
\eqno(4.23)
$$
To give this equation in the same parametrisation as (4.3), we first write
it in terms of $\yup$ and then use a
multiplicative gauge
$\omega=\phi w$ to put to unity the coefficient of $\omega y$ in the
numerator of $\overline \omega$. We then have:
$$
\overline{\omega}={\omega(\yup-r)+q(\yup-\overline{\psi}_3) \over
\omega\yup}. \eqno(4.24)
$$
where
$$
\matrix{
\displaystyle{q={(p-\pup)(\pdo-\pdd)(2\cdd+\cdo+1) \over
(\cdo(p-\pdo)+\pup-\pdo)(\cdd(\pdo-\pdd)+p-\pdd)}}\cr \cr
\displaystyle{r=
{\cdo(p-\pdo)+c(\pup-\pdo) \over \cdo(p-\pdo)+\pup-\pdo}} \cr \cr
\displaystyle{
\overline{\psi}_3={\cdd c+c+\cdo+\cdd \over 2\cdd+\cdo+1}.}
}
\eqno(4.25)
$$

Finally, we examine the possibility of the existence of a $\Delta N=1$
Schlesinger. In this case, the
structure of the transformation will be obtained by asking that the $N+1$
case enter the singularity
one step before the $N$ case but exit at the same point. The general
structure is thus:
$$
w={rx+q\psi_N \over y}{y-\eta \over x-\xi} \eqno(4.26)
$$
where $\eta$ and $\xi$ must be determined. We do this by requiring that the
equation for $w$ contain no
coefficients nonlinear in $y$. As a result we find that $\eta$ must satisfy
the equation (4.1) for $y$:
$$
\overline{\eta}={\eta+c \over \eta+1} \eqno(4.27)
$$
and $\xi$ the equation (4.3) for $x$ with $\eta$ instead of $y$:
$$
\overline{\xi}={\xi(\eta-r)+q(\eta-\psi_N) \over \xi \eta}. \eqno(4.28)
$$
We remark here the perfect parallel to the continuous case (and as we
pointed out the discrete case led
the investigation back to the continuous one). Let us point out here that
the $w$ obtained through
(4.26) does not lead to $w=0$ at the exit of the singularity (i.e. when
$x=0$, $y=\psi_N$) and a translation is needed.
One has in principle to define a new variable
$$
\omega = w-w(x=0,y=\psi_N)=w+{q \over \xi}(\psi_N-\eta).
$$

We are now going to study the particular case where we construct $N=1$
starting from $N=0$. As in the continuous case our
starting point is the $N=0$ nonlinear equation. Thus we are starting with
the decoupled equation (4.14) for $x$.
We write the Schlesinger as:
$$
w={y-\eta \over y}{x \over x-\xi}. \eqno(4.29)
$$
Once more we find that $\eta$ must satisfy (4.27) while $\xi$ satisfies
$\overline{\xi}=1+q / \xi$ i.e. the same
equation as $x$. Using (4.27) we can easily obtain the equation for $w$
corresponding to $N=1$. We write this equation in
terms of $\yup$ in order to enter the singularity with $\yup=0$. The
expression of $\overline w$ is indeed linear in $w$
and reads
$$
\overline{w}={(q+\xi)w(c-\yup)+q(\yup(\eta+1)-c-\eta) \over \yup q (\eta+1)
}. \eqno(4.30)
$$
In order to cast it in the parametrisation of (4.21), we introduce a shift
in the $w$:
$$
\omega=w+\omega_0. \eqno(4.31)
$$
We require that in the numerator of the equation for $\omega$, $\yup$
appears only as a product with $\omega$. We find
that
$\omega_0$ must satisfy
$$
\overline{\omega}_0=-{\omega_0(q+\xi)+q(\eta+1) \over q(\eta+1)} \eqno(4.32)
$$
and we obtain the following equation for $\omega$
$$
\overline{\omega}={(q+\xi)\omega(c-\yup)-c\omega_0(q+\xi)-q(c+\eta) \over
\yup q(\eta+1)} \eqno(4.33)
$$
which is in the form of (4.21) up to a multiplicative factor in $\omega$.
We note that $\omega_0$ plays the role of
$\tilde{x}_0$ in equation (3.18) in the continuous case. Indeed, $\omega_0$
does satisfy (4.33) for $y=\eta$.

Finally we derive the $\Delta N=1$ Schlesinger for the case of a linear
mapping (4.18). We start from:
$$
w={x\psi_N-g \over y}{y-\eta \over x-\xi} \eqno(4.34)
$$
and again require for $w$ an equation with coefficients linear in $y$. We
find that $\eta$ must again be
a solution of the equation for $y$ i.e. it must satisfy (4.27) and moreover
$\xi$ is a solution of
(4.18) with $y=\eta$:
$$
\overline{\xi}={\xi(\eta-\psi_N)+g \over \eta}. \eqno(4.35)
$$
Thus the list of the Schlesinger transformations of the Gambier mapping is
complete.

\bigskip
\noindent {\scap 5. Conclusion}
\medskip

In this paper we have shown that the Gambier system possesses Schlesinger
transformations just like the
Painlev\'e equations. This is a most interesting result given that the
Gambier system is C-integrable (in
the terminology of Calogero [10]) i.e. integrable through linearisation,
and not S-integrable.

We discovered that both the continuous and the discrete systems possess two
kinds of Schlesinger
transformations: one that allows changes of $N$ by two units and one where
the changes of $N$ are by one
unit. In the discrete case our approach was based entirely on the
singularity confinement aproach. We
have shown that a study of the singularities allows one to determine the
form of the Gambier mapping and
at the same time its Schlesinger transformations. This is one more argument
in favour of the singularity
analysis approach to the study of discrete systems.

 \bigskip
\noindent {\scap Acknowledgements}.
\smallskip
\noindent
 S. Lafortune acknowledges two
scholarships: one from NSERC (National Science and Engineering Research
Council of Canada) for his Ph.D.
and one from ``Programme de Soutien de Cotutelle de Th\`ese de doctorat du
Gouvernement du Qu\'ebec'' for
his stay in Paris.
\bigskip
{\scap References}
\smallskip
\item{[1]} M. Jimbo and T. Miwa, Physica D2 (1981) 407, D4 (1981) 47.
\item{[2]} A.S. Fokas and M.J. Ablowitz,
 J.Math.Phys. 23 (1982) 2033.
\item{[3]} B. Grammaticos, F. Nijhoff and A. Ramani, {\sl Discrete
Painlev\'e equations},
course at the Carg\`ese 96 summer school on Painlev\'e equations.
\item{[4]}	A. Ramani, B. Grammaticos and J. Satsuma,
Jour. Phys. A 28 (1995) 4655.
\item{[5]} N.Joshi, A.Ramani and B.Grammaticos, {\sl A Bilinear Approach to
Discrete Miura Transformations}, preprint 98.
\item{[6]} B. Gambier, Acta Math. 33 (1910) 1.
\item{[7]} B. Grammaticos and A. Ramani, Physica A 223 (1995)
125.
\item{[8]} B. Grammaticos, A. Ramani and S. Lafortune, {\sl The Gambier
mapping,
revisited}, to appear in Physica A.
\item{[9]} E.L. Ince, {\sl Ordinary differential equations}, Dover, New
York, 1956.
\item{[10]} F. Calogero, {\sl Why Are Certain Nonlinear PDEs Both Widely
Applicable and Integrable?}, in {\sl What is
Integrability?}, V.E. Zakharov (Ed.), 1-62, Springer-Verlag, Berlin, 1991.
\end